\documentclass[11pt,onecolumn,amssymb,nofootinbib]{revtex4}
\usepackage{amsmath, amsthm, amscd, amssymb}
\usepackage{bm}
\usepackage{bbm}

\begin{document}

\title{\bf Implications of a frame dependent dark energy for the spacetime metric, cosmography, and effective Hubble constant}

\author{Stephen L. Adler}
\email{adler@ias.edu} \affiliation{Institute for Advanced Study,
Einstein Drive, Princeton, NJ 08540, USA.}

\begin{abstract}
In earlier papers we showed that a frame dependent effective action motivated by the postulates of three-space
general coordinate invariance and Weyl scaling invariance exactly mimics a cosmological constant in Friedmann-Robertson-Walker (FRW)
spacetimes, but alters the linearized equations governing scalar perturbations around a spatially flat FRW background metric.   Here we analyze the implications of a frame dependent dark energy for the
spacetime cosmological metric within both a perturbative and a non-perturbative framework.  Both methods of calculation give a one-parameter family of cosmologies which are in close correspondence to one another, and which contain the standard FRW cosmology as a special case. We discuss the application of this family of cosmologies to the standard cosmological distance measures and to the effective Hubble parameter,
with special attention to the current tension between determinations of the Hubble constant at late time, and the Hubble value obtained through the cosmic microwave background (CMB) angular fluctuation analysis.

\end{abstract}

\maketitle

\section{Introduction}

The experimental observation of an accelerated expansion of the universe has been interpreted as evidence for a cosmological term in the gravitational action of the usual form
\begin{equation}\label{usual}
 S_{\rm cosm}=-\frac{\Lambda}{8 \pi G} \int d^4x (^{(4)}g)^{1/2}~~~,
\end{equation}
with $\Lambda=3H_0^2\Omega_{\Lambda}$  in terms of the Hubble constant $H_0$ and the cosmological fraction $\Omega_{\Lambda}$.
This functional form incorporates the usual assumption that gravitational physics is four-space general coordinate invariant,
with no frame dependence in the fundamental action.

In a series of papers \cite{adler1}-\cite{adler4}, motivated by the frame dependence of the CMB radiation, and ideas about scale invariance of an underlying pre-quantum theory,  we have studied the implications of the assumption that there is an induced
gravitational effective action that is three-space general coordinate and Weyl scaling invariant, but is not four-space general coordinate invariant.  This analysis leads to an alternative dark energy action given by \begin{equation}\label{effact3}
 S_{\rm eff}=-\frac{\Lambda}{8\pi G} \int d^4x (^{(4)}g)^{1/2} (g_{00})^{-2}~~~,
\end{equation}
which in Friedmann-Robertson-Walker (FRW) spacetimes where $g_{00}=1$ exactly mimics the cosmological constant effective action of Eq. \eqref{usual}.

To set up a phenomenology for testing for the difference between standard and frame-dependent dark energy actions, we made
the ansatz that the observed cosmological constant arises from a linear combination of the two  of the form
\begin{equation} \label{ansatz}
S_{\Lambda} = (1-f)  S_{\rm cosm}+ f  S_{\rm eff}
= -\frac{\Lambda}{8\pi G} \int d^4x (^{(4)}g)^{1/2} [1-f + f(g_{00})^{-2}]~~~~,
\end{equation}
so that $f=0$ corresponds to only a standard cosmological constant, and $f=1$ corresponds to only an apparent cosmological constant
arising from a frame dependent effective action. In \cite{adler4} we gave detailed results for the scalar
perturbation equations around a FRW background arising when dark energy is included through the
action of Eq. \eqref{ansatz}.  Our aim in this paper is to analyze the implications of the ansatz of
Eq. \eqref{ansatz} for the spacetime cosmological metric, with applications to standard cosmological
distance measures and the effective Hubble parameter. We will carry out the calculations in two ways,
first by using the linearized perturbation equations around a FRW background derived in \cite{adler4},
and then by a non-perturbative method, giving  results that are in close
correspondence.  Both methods yield a one-parameter family of cosmologies, parameterized by the initial
value at cosmic time $t=0$ of one of the metric components, with the standard FRW cosmology included as
a special case.

This paper is organized as follows.\footnote{An earlier version \cite{adler5} has been merged into this one.}     In Sec. 2 we state our metric in non-perturbative and perturbative
forms, and explain why when $f\neq 0$ the model of Eq. \eqref{ansatz} cannot be reduced to a standard
cosmological constant by redefinition of the time variable.  We also introduce some notations that
will be used frequently later on.  In Sec. 3 we solve for the metric dynamics implied by Eq. \eqref{ansatz} using the perturbative formalism.  In Sec. 4 we repeat this calculation by a non-perturbative method
and show that the results closely correspond to those of Sec. 3.  In Sec. 5 we give results for the
standard cosmographic distance measures and the effective Hubble parameter, and apply
our results to the recently much discussed ``Hubble tension''.  In  Appendix A  we give details of the
perturbative derivation of the metric dynamical equation, which is also obtained as a by-product of the
non-perturbative analysis.  In Appendix B we give formulas for comparing our model with results from baryon acoustic oscillation (BAO) measurements.

\section{Homogeneous isotropic metric, why the $f\neq 0$ case does not reduce to $f=0$, and some
notations}

Our starting point is the observation that the general form for the line element in a homogeneous, isotropic,
zero spatial curvature universe in which physics is invariant under three-space general coordinate
transformations, but not invariant under four-space general coordinate transformations, is
\begin{equation}\label{metric}
ds^2=\alpha^2(t) dt^2-\psi^2(t) d\vec x^{\,2}=g_{00} dt^2+ g_{ij}dx^idx^j~~~,
\end{equation}
with metric components
\begin{align}\label{metriccomps}
g_{00}=&\alpha^2(t)~~,~~~g_{ij}=-\delta_{ij}\psi^2(t)~~~,\cr
g^{00}=&1/\alpha^2(t)~~,~~~g^{ij}=-\delta_{ij/}\psi^2(t)~~~.\cr
\end{align}

\subsection{Why the $f \neq 0$ case cannot be reduced to $f=0$ by redefining the time variable}

An evident feature of Eq. \eqref{metric} is that if we define a proper time $\tau$ by\footnote{We have
chosen the arbitrary lower limit of the integral in Eq. \eqref{propertime} so that the proper time
origin $\tau=0$ coincides with the coordinate time origin $t=0$.}
\begin{equation}\label{propertime}
d\tau=\alpha(t)dt~~,~~~\tau=\int_0^t du \alpha(u)~~~,
\end{equation}
with inversion $t(\tau)$,
and denote $\psi(t)$ as written in terms of $\tau$ by $\psi(t(\tau))=\psi[\tau]$,
then Eq. \eqref{metric} takes
the form
\begin{equation}\label{metric1}
ds^2=d\tau^2-\psi^2[\tau]d\vec x^{\,2}~~~.
\end{equation}
This has the same form as the standard FRW metric
\begin{equation}\label{frwmetric}
ds^2=dt^2-a^2(t)d\vec x^{\,2}~~~,
\end{equation}
with $\tau$ replacing $t$ and with $\psi(t)=\psi[\tau]$ replacing the FRW expansion factor $a(t)$.
However, this does not mean that Eq. \eqref{ansatz} with $f\neq 0$ can be reduced to the $f=0$ case.
To see this, we rewrite Eq. \eqref{ansatz} in terms of the metric components $\alpha(t)$ and
$\psi(t)$, both assumed positive so that $(^{(4)}g)^{1/2}=\alpha(t)\psi^3(t)$, giving
\begin{equation} \label{ansatz1}
S_{\Lambda}= -\frac{\Lambda}{8\pi G} \int dt  d^3x \alpha(t) \psi^3(t)[1-f + f\alpha(t)^{-4}]
= -\frac{\Lambda}{8\pi G}
\int d\tau  d^3x\psi^3[\tau] [1-f + f\alpha[\tau]^{-4}]~~~.
\end{equation}
When $f=0$ the metric component $\alpha(t)$ is absorbed into the definition of the new time variable, as
expected since the standard cosmological action is four-space general coordinate invariant.  But when
$f\neq 0$, the factor $\alpha[\tau]^{-4}$ cannot be similarly absorbed, reflecting the fact that the $f$
term is only three-space, but not four-space general coordinated invariant.\footnote{If one were to define $d\tau=dt\alpha(t)[1-f+f/\alpha(t)^4]$, the Einstein-Hilbert action which is proportional to $ \int dt d^3x \alpha(t)\psi^3(t) R$ would be left with a residual $\alpha(t)$ dependence.  The same is true if one were to define $d\tau=\alpha(0) dt$.}  So we anticipate that
Eq. \eqref{ansatz} with $f\neq 0$ will give a more general cosmology than the standard FRW cosmology.
However, since when $\alpha(t) \equiv 1$ Eq. \eqref{ansatz} reduces back to a standard cosmological term,
we also anticipate that this more general dynamics will include the standard FRW cosmology as a special case.

\subsection{Some notations}

We record the following notational conventions that will be used throughout.

\begin{itemize}

\item  The metric components $\alpha(t)$ and $\psi(t)$ form the basis of our non-perturbative
treatment.  But in both the non-perturbative and perturbative discussions, we will also use the related  components $\theta(t)$, $\Phi(t)$ and $\Psi(t)$ defined by\footnote{The factor $1/\theta(0)$ drops out
of the equations below from which we calculate $\psi(t)$, since these are homogeneous  in $\psi$. Similarly, it is not relevant for the equations of Sec. 3 where we calculate $\Phi(t)$ and $\Psi(t)$, since these equations are homogeneous in $a(t)$.  But it guarantees that at early times $\psi(t)$ becomes identical to $a(t)$, which is important for the match
to the CMB fits, since these are based on small fluctuation equations including $\vec x$ dependence that are sensitive to the absolute magnitude of $a(t)$.}
\begin{align}\label{PhiPsi}
\alpha(t)=&1+\Phi(t)~~~,\cr
\psi(t)=&a(t)\theta(t)/\theta(0)~~~,\cr
\theta(t)=&1-\Psi(t)~~~,\cr
\end{align}
with $a(t)$ the standard FRW expansion factor.

\item  Since our focus is on the matter-dominated era, we use the following convenient approximate
formula  \cite{kolb} for $a(t)$,\footnote{Before the matter-dominated era,
when the radiation content of the universe is significant, $a(t)$ is no longer well approximated by
Eq. \eqref{closed}.  But this does not affect the numerical results given below.}

\begin{equation}\label{closed}
a(t)\simeq \left(\frac{\Omega_m}{\Omega_{\Lambda}}\right)^{1/3} \big(\sinh(x)\big)^{2/3}~~~,~~\Omega_m=1-\Omega_{\Lambda}~~~,
\end{equation}
with $x$ the dimensionless time variable
\begin{equation}\label{dimtime}
x=\frac{3}{2} \surd{\Omega_{\Lambda}} H_0 t~~~.
\end{equation}
At the present era $t=t_0$ defined by $a(t_0)=1$, $x_0$ takes the value
\begin{equation}\label{dimtime0}
x_0= {\rm arcsinh}((\Omega_\Lambda/\Omega_m)^{1/2}) \simeq 1.169 ~~~.
\end{equation}
Eqs. \eqref{dimtime} and \eqref{dimtime0} then give  $H_0t_0=(2/3)x_0/\surd{\Omega_\Lambda}=0.946$, in
agreement with the Planck values for this product,
where we have used the Planck 2018 values \cite{planck} $\Omega_{\Lambda}=0.679$, $\Omega_m=0.321$.
However, we will see below in Sec. 5 that $H_0$ in Eqs. \eqref{closed} and \eqref{dimtime} are rescaled by factor $\alpha(0)$ from the Planck value $H_0^{\rm Pl}\simeq 67{\rm km} {\rm s}^{-1} {\rm Mpc}^{-1}$, so that $H_0=\alpha(0)H_0^{\rm Pl}$.
\item
The Hubble parameter $H(t)$ defined by the standard FRW cosmology is
\begin{equation}\label{frwH}
H(t)= \frac{da(t)/dt}{a(t)}=\frac{\dot a(t)}{a(t)}~~~,
\end{equation}
which in the matter-dominated era, using the approximate formula of Eq. \eqref{closed}, is
\begin{align}\label{hoft}
H(t)=&H_0\surd{\Omega_{\Lambda}} \coth(x)~~~\cr
=&H_0[\Omega_m(1+z)^3+\Omega_{\Lambda}]^{1/2}~~~,\cr
\end{align}
with $\coth(x)=\cosh(x)/\sinh(x)$ and
with the redshift $z$ defined by $1+z=1/a(t)$.
This is to be distinguished from the Hubble parameter $H_{\rm eff}(t)$ arising from the modified
dynamics of Eq. \eqref{ansatz}, which is given by the proper time derivative
\begin{equation}
H_{\rm eff}[\tau]= \frac{d\psi[\tau]/d\tau}{\psi[\tau]}=H_{\rm eff}(t)=
\frac{d\psi(t)/dt}{\alpha(t)\psi(t)}~~~.
\end{equation}
\end{itemize}

\section{Metric perturbation derivation}

\subsection{Setting up the $\Phi$ equation}

In \cite{adler4} we gave detailed results for the scalar
perturbation equations around a FRW background arising when dark energy is included through the
action of Eq. \eqref{ansatz}.  Writing \cite{wein}\footnote{The equations of Sec. 3 and Appendix A follow  \cite{wein} and \cite{adler4} and do not
include the factor $1/\theta(0)$ of Eq. \eqref{PhiPsi}, which can be thought of here as residing in the normalization of $a$.   This factor is not relevant for the analysis of this section since the equations that we use to determine $\Phi$ and $\Psi$ are homogeneous in $a$, and so
the normalization of $a$ drops out.}
\begin{align}\label{split}
g_{00}=&1+E~~~,\cr
g_{i0}=&-a(t)(\partial_iF+{\rm vector})~~~,\cr
g_{ij}=&-a^2(t)[(1+A)\delta_{ij}+\partial_i\partial_jB+{\rm vector}+{\rm tensor}]~~~,
\end{align}
we gave formulas in $B=0$ gauge for the linearized equations governing the scalar perturbations $A$,  $E$, and $F$, for
the general case in which these are functions of both space and time. For the analysis of this paper,
it is more convenient to express the $B=0$ gauge linear perturbation equations  in terms of $F$ and the functions $\Phi$ and $\Psi$ that are related to  $E$ and $A$ by
\begin{align}\label{EAelim}
E=&2[\Phi-\partial_t(aF)]~~~,\cr
A=&-2[\Psi+\dot a F]~~~,\cr
\end{align}
as given in Appendix A.
This form of the perturbation equations corresponds to writing the perturbed line element as
\begin{equation}\label{line}
ds^2=[1+2\Phi]dt^2-a^2(t)[1-2\Psi] d\vec x^2~~~,
\end{equation}
in agreement to linear order with Eqs. \eqref{metric} and \eqref{PhiPsi} above when the $1/\theta(0)$ factor is omitted (or is  included in the normalization of the expansion factor $a$).

In the limit that the metric perturbations have no spatial dependence, the perturbation $F$, which
appears in Eq. \eqref{split} acted on by spatial derivatives, can be assumed to have the $\vec x$-independent limit $F(t)=0$  and so can be dropped from the metric perturbation
equations.\footnote{This still allows one to make three-space general coordinate transformations, but not
four-space general coordinate transformations, which corresponds to the invariance properties of the
action of Eq. \eqref{ansatz}.}   Also, in the matter-dominated era the anisotropic inertia \cite{wein} $\pi^S=0$, so assuming continuity
in the limit of vanishing spatial dependence, the $Y=0$ part of Eq. \eqref{ijeq} in Appendix A implies that $\Phi(t)=\Psi(t)$.  With these simplifications, we show in Appendix A that
 the perturbation equations governing the time evolution of $\Phi(t)$ can be put into the form (with  $\dot{}=d/dt$)
\begin{equation}\label{evolution1}
\ddot \Phi+4\frac{\dot a}{a} \dot \Phi + \left(2 \frac{\ddot a}{a}+\frac{\dot a^2}{a^2}\right)\Phi =
4\pi G \delta p + 2\Lambda f \Phi~~~,
\end{equation}
with $\delta p$ the pressure perturbation. As noted, this equation is independent of the normalization
of $a(t)$.   When the term proportional to $f$ is dropped, this agrees with Eq. (7.49) of Mukhanov \cite{muk} and Eq. (23c) of Ma and Bertschinger \cite{ma} when their conformal
time derivatives are converted to time derivatives.
In the matter-dominated era, $\delta p$ vanishes for adiabatic perturbations, so then  we can drop the $\delta p$ term in Eq. \eqref{evolution1}, giving
\begin{equation}\label{evolution11}
\ddot \Phi+4\frac{\dot a}{a} \dot \Phi + \left(2 \frac{\ddot a}{a}+\frac{\dot a^2}{a^2}\right)\Phi =
 2\Lambda f \Phi~~~.
\end{equation}
It is convenient now  to use the dimensionless time variable $x$ introduced above in Eq.  \eqref{dimtime}, which when substituted into Eq. \eqref{evolution11} gives finally the evolution equation for $\Phi$ in terms of $x$,
\begin{equation}\label{evolution2}
\frac{d^2 \Phi}{dx^2}+\frac{8}{3} \coth(x) \frac{d \Phi}{dx} =\frac{4}{3}(2f-1) \Phi~~~.
\end{equation}

\subsection{Large and small $x$ behavior and numerical solution }

Before proceeding to the numerical solution of Eq. \eqref{evolution2}, we examine analytically the large and small $x$ behavior of the solutions.  For large $x$ the function $\coth(x)$ approaches unity, and
Eq. \eqref{evolution2} becomes an equation with constant coefficients solved by exponentials,
\begin{align}\label{expsol}
\Phi(x)=&C_1e^{\mu_+ x}+C_2 e^{\mu_- x}~~~,\cr
\mu_{\pm}=&-\frac{2}{3}[2 \pm (6f+1)^{1/2}]~~~.\cr
\end{align}
As suggested already by the factor $2f-1$ in Eq. \eqref{evolution2}, there is a crossover in behavior
at $f=1/2$. For $f<1/2$, both exponents in Eq. \eqref{expsol} are negative, and $\Phi$ decays to zero as
$x\to \infty$.  For $f=1/2$, one exponent is negative, while the other is zero, and $\Phi$ approaches a
constant as $x\to \infty$.  For $f>1/2$, one exponent  remains negative, while one is positive, and $\Phi$ becomes infinite as $x \to \infty$.  So for $f=1$, the case of a scale invariant cosmological action, the
metric perturbation $\Phi$ grows with time.

We examine next the small $x$ behavior, where the term proportional to $2f-1$ becomes much less important
than the terms on the left of Eq. \eqref{evolution2}.  This equation is then approximated by
\begin{equation}\label{evolution3}
\frac{d^2 \Phi}{dx^2}+\frac{8}{3} \frac{1}{x} \frac{d \Phi}{dx} =0~~~,
\end{equation}
with the general solution
\begin{equation}\label{smallx}
\Phi(x)=C_3+C_4x^{-5/3}~~~.
\end{equation}
This shows that sufficient boundary conditions for getting a unique solution are the requirements that
the solution be regular at $x=0$, together with specification of the value $\Phi(0)$.

With this analysis in mind, we rewrite Eq. \eqref{evolution2} as an integral equation.  Defining the normalized perturbation
\begin{equation}\label{phihat}
\hat \Phi(x)=\Phi(x)/\Phi(0)~~~,
\end{equation}
 $\hat \Phi$ obeys the integral equation
\begin{equation}\label{inteq}
\hat \Phi(x)=1+\frac{4}{3}(2f-1) \int_0^x dw (\sinh(w))^{-8/3} \int_0^w du (\sinh(u))^{8/3}\hat\Phi(u)~~~,
\end{equation}
which incorporates both the boundary conditions at $x=0$  and the differential equation of
Eq. \eqref{evolution2}.  Starting from an initial assumption $\hat \Phi(x)=1$, and then updating at each
evaluation of the right hand side of Eq. \eqref{inteq}, the integral equation converges rapidly to
an answer in 5 iterations on an 800 bin mesh\footnote{ On this coarse mesh, the  value $x_{\rm eq}<10^{-5}$ at the transition from radiation domination to  matter domination is indistinguishable from 0, so it suffices to take zero as the lower limit of the integrals.}
 taking integrand values at center-of-bin.  The results for $d\hat \Phi/dx$  with $f=1$ are given in Table I,  for various values of $x$ ranging from 0 to $x_0$, showing that $d\hat \Phi/dx|_{x=x_0}$ is positive.  Consistent with the large time analysis given above, when we repeat the calculation with $f=0$ we find a negative value of  $d\hat \Phi/dx|_{x=x_0}$.  In the final column of Table I we tabulate the quadratic
 approximation to $\hat \Phi(x)$,
 \begin{equation}\label{quadapprox}
 \hat \Phi(x)\simeq 1 + C (x/x_0)^2~~,~~~C=0.244~~~,
 \end{equation}
 which is accurate to a few parts per thousand over the entire interval $0 \leq x/x_0 \leq 1$.

\begin{table} [ht]
\caption{Values of $d\hat \Phi/dx$ and $\hat \Phi$  versus $x/x_0$ and redshift $z$, calculated with $f=1$, that is, with all of
the observed cosmological constant arising from a scale invariant but frame dependent action.  The final
column gives the fit of $\hat \Phi$ to the quadratic $1+C(x/x_0)^2$, with $C=0.244$.}
\centering
\begin{tabular}{c  c c c c}
\hline\hline
$x/x_0$ & $z$& $d\hat \Phi/dx$  & $\hat \Phi$   &   $1+0.244(x/x_0)^2$\\
\hline
1.0 & 0.00       &   0.409   & 1.244   &   1.244     \\
0.9 & 0.10       &   0.369   & 1.198   &   1.197    \\
0.8 & 0.22      &   0.331   & 1.158    & 1.156       \\
0.7 & 0.36        &  $ 0.290 $  &1.121 &  1.120      \\
0.6 & 0.54      &  $ 0.251 $  &1.089   & 1.088    \\
0.5 & 0.77       &   $ 0.209 $  &1.062 & 1.061       \\
0.4 & 1.08      &   $ 0.169 $ &1.040   & 1.039      \\
0.3 & 1.55       &   $ 0.126  $ &1.023 & 1.022        \\
0.2 & 2.36       &   $ 0.0848 $  &1.010 & 1.0098       \\
0.1 & 4.36      &    $ 0.0416  $  &1.0026 & 1.0024    \\
0.0341 & 10 &  0.0131  &  1.0003 & 1.000028  \\
0.00122  &100  &  0.000018  &  1.0000 & 1.0000\\
\hline\hline
\end{tabular}
\label{tab1}
\end{table}

\section{Non-perturbative derivation}

\subsection{ Einstein tensor and equations, and the dark energy and matter energy momentum tensors}

We return now to the general form for the line element given in Sec. 2,
\begin{equation}\label{metricaa}
ds^2=\alpha^2(t) dt^2-\psi^2(t) d\vec x^{\,2}~~~.
\end{equation}
Using a Mathematica notebook for general relativity \cite{mathematica}, the nonzero Einstein tensor components for this metric are computed to be
\begin{align}\label{Einstein}
G_{00}=&-3\frac{\dot \psi^2(t)}{\psi^2(t)}~~~,\cr
G_{ij}=&\delta_{ij}\frac{2\psi(t)[\alpha(t)\ddot \psi(t)-\dot \alpha(t)\dot \psi(t)]+ \alpha(t) \dot \psi^2(t)}{\alpha^3(t)}~~~\cr
\end{align}
where $\dot \psi^2(t)=(d\psi(t)/dt)^2$.

Writing  the dark energy action of Eq. \eqref{ansatz} as
\begin{equation} \label{ansatz11}
S_{\Lambda}
= -\frac{\Lambda}{8\pi G} \int d^4x (^{(4)}g)^{1/2} [1-f + f/\alpha^4(t)]~~~~,
\end{equation}
and  varying this equation with respect to $g_{ij}$,
we get the $ij$  component of the dark energy contribution to the energy momentum tensor,
\begin{equation}\label{tlambda}
T_{\Lambda}^{ij}=\frac{\Lambda}{8\pi G}[(1-f)g^{ij}+f t^{ij}]~,~~
T_{\Lambda \,ij}=\frac{\Lambda}{8\pi G}[(1-f)g_{ij}+f t_{ij}]~~~,
\end{equation}
with
\begin{equation}\label{tcomps}
t^{ij}=-\delta_{ij}\frac{1}{\alpha^4(t) \psi^2(t)}~~,~~~
t_{ij}=-\delta_{ij}\frac{\psi^2(t)}{\alpha^4(t)}~~~.
\end{equation}
As detailed in \cite{adler1}--\cite{adler4}, in order to use this as the source term in the
Einstein equations, the remaining components of $t_{\mu\nu}$ must be determined so that covariant
conservation with respect to the metric is satisfied.  The conserving completion of the $g_{ij}$ part
of Eq. \eqref{tlambda} is just $g_{\mu\nu}$.  Since the metric is diagonal, the conserving completion
of $t_{ij}$ in  Eq. \eqref{tcomps} has $t_{0i}=t_{i0}=0$, and $t_{00}$ given by the solution
of the covariant conservation equation, which is
\begin{equation}\label{covcons}
\dot t_{00}(t) + t_{00}(t) \left[3 \frac{\dot \psi(t)}{\psi(t)}-2\frac{\dot \alpha(t)}{\alpha(t)}\right]
=3 \frac{\dot \psi(t)}{\alpha^2(t) \psi(t)}~~~.
\end{equation}
This equation can be readily integrated to give
\begin{equation}\label{covonsint}
t_{00}(t)=3 \frac{\alpha^2(t)}{\psi^3(t)}\int_0^t du \frac{\dot \psi(u) \psi^2(u)}{\alpha^4(u)}~~~,
\end{equation}
where we have arbitrarily taken the lower limit of integration as zero (more about this below).

We additionally need the particulate matter energy momentum tensor, for which we take the relativistic   perfect fluid form
\begin{equation}\label{partm}
T_{\mu\nu}^{\rm pm}=(p+\rho)u_{\mu}u_{\nu}-pg_{\mu\nu}~~~,
\end{equation}
with $p$ the pressure, $\rho$ the energy density, and $u_{\mu}=g_{\mu\nu}dx^{\nu}/ds$ the four velocity.
For the metric of Eq. \eqref{metric}, $u_0$ is given by
\begin{equation}\label{uzero}
u_0=\alpha^2(t) \frac{dx^0}{ds}=\alpha^2(t) \frac{dx^0}{\alpha(t) dt}
=\alpha(t)\frac{dx^0}{dt}=\alpha(t)~~~,
\end{equation}
from which we find
\begin{align}\label{tpmcomps}
T_{00}^{\rm pm}=& \rho \alpha^2(t) ~~~,\cr
T_{ij}^{\rm pm}=& \delta_{ij} p \psi^2(t)~~~.\cr
\end{align}
Covariant conservation of $T_{\mu\nu}^{\rm pm}$ implies
\begin{equation}\label{cons1}
\frac{d\big(\rho(t)\psi^3(t)\big)}{dt}= -3 p(t) \dot \psi(t) \psi^2(t)~~~,
\end{equation}
which dividing by $\alpha(t)$ on both sides is equivalent to
\begin{equation}\label{cons2}
\frac{d\big(\rho[\tau]\psi^3[\tau]\big)}{d\tau}= -3 p[\tau] \frac{d\psi[\tau]}{d\tau} \psi^2[\tau\emph{}]~~~.
\end{equation}

\subsection{Einstein equations}

We now have all the ingredients needed to write down the Einstein equations. From the $00$ component we get
\begin{equation}\label{00comp}
\frac{\dot \psi^2(t)}{\alpha^2(t)\psi^2(t)}=\frac{\Lambda}{3}\left[ 1-f+f
\frac{3}{\psi^3(t)} \int_0^t du \frac{\dot \psi(u) \psi^2(u)}{\alpha^4(u)}\right] +
\frac{8\pi G}{3}\rho(t)~~~,
\end{equation}
and from the spatial components we get
\begin{equation}\label{spatcomp}
\frac{2 \psi(t)[\alpha(t) \ddot\psi(t)-\dot\alpha(t)\dot\psi(t)] +\alpha(t) \dot \psi^2(t)}{\alpha^3(t)}
=\psi^2(t)\left[(1-f)\Lambda +\frac{f \Lambda}{\alpha^4(t)}-8 \pi G p(t) \right]~~~.
\end{equation}
As noted above, both of these equations are independent of the normalization of $\psi(t)$.

The three equations Eq. \eqref{00comp}, Eq. \eqref{spatcomp}, and Eq. \eqref{cons1} are not
independent, by virtue of the Bianchi identities for $G_{\mu\nu}$ and covariant conservation of
the energy momentum tensors.  Multiplying Eq. \eqref{00comp} through by $\psi^3(t)$, differentiating
with respect to time,  eliminating $(d/dt)(\psi^3(t) \rho(t))$ by Eq. \eqref{cons1}, and dividing by
 $\dot \psi(t)$, we recover
Eq. \eqref{spatcomp}.  Hence effectively we  have only one equation relating the metric components
$\alpha$ and $\psi$, and a second equation will be needed, a topic which we address in the next section.

But first let us address several issues raised by the structure of the above equations, beginning with
the arbitrary lower limit in the integral in Eq. \eqref{covonsint}.
In the sequel we will be interested  in solving Eqs. \eqref{00comp} and \eqref{spatcomp} in the
matter-dominated era, in which the pressure
$p=0$.  Then Eq. \eqref{cons1} can be integrated to give $\rho(t)=C/\psi^3(t)$, with $C$ a constant of
integration, and we see that a shift in the lower limit of the integral in Eqs. \eqref{covonsint} and \eqref{00comp}  can be absorbed into a shift in the integration constant $C$.  So there is no loss of  generality in
taking the lower limit of integration as zero.\footnote{When Eq. \eqref{00comp} is converted to Eq. \eqref{spatcomp} by multiplication by $\psi^3(t)$  and differentiation, the constant $C$ drops out, and is replaced by the additional constant of integration needed for a second order differential equation as opposed to a first order equation.    This new
constant will end up parameterizing the family of solutions of our model.}  A further property of Eqs. \eqref{00comp} and \eqref{spatcomp} is that when $\alpha(t)\equiv 1$, they reduce when $\psi(0)=0$ to the usual FRW equations with $\psi(t)$ playing the role of the FRW expansion factor $a(t)$,
\begin{align}\label{alpha1}
\frac{\dot \psi^2(t)}{\psi^2(t)}=&\frac{\Lambda}{3} +\frac{8\pi G}{3}\rho(t)~~~,\cr
2 \psi(t) \ddot\psi(t) + \dot \psi^2(t)=&\psi^2(t)[\Lambda -8 \pi G p(t) ]~~~.\cr
\end{align}
These equations have as the  solution $\psi(t)=a(t)$,
a property that will be relevant for finding a second equation relating $\alpha(t)$ and $\psi(t)$.

\subsection{Finding the second equation}

To find a second equation relating $\alpha(t)$ and $\psi(t)$, we generalize to the case where these
are also functions of the coordinate $\vec x$, so that we have $\alpha(t,\vec x)\,,\,\psi(t,\vec x)$,
and then assume a smooth limit as the coordinate dependence limits to zero.  When $\alpha$ and $\psi$ are
coordinate dependent, the Einstein tensor component $G_{xy}$ is given by
\begin{equation}\label{gxy}
\psi^2\alpha G_{xy}=-2\alpha \partial_x \psi \partial_y\psi+ \alpha \psi \partial_x\partial_y \psi
+\psi^2\partial_x\partial_y\alpha-\psi(\partial_x\alpha \partial_y \psi+\partial_y\alpha \partial_x
\psi)~~~,
\end{equation}
which is independent of the normalization of $\psi(t,\vec x)$.
Since the energy momentum source tensors have no $xy$ component, $G_{xy}$ must be equated to zero.
To proceed further, with no loss of generality so far we write $\alpha$ and $\psi$ as
\begin{align}\label{alphapsi}
\alpha(t,\vec x)=&1+\Phi(t,\vec x) ~~~,\cr
\psi(t,\vec x)=&a(t) \, \theta(t, \vec x)/\theta(0,\vec 0) ~~~,\cr
\theta(t,\vec x) =&1-\Psi(t, \vec x)~~~,\cr
\end{align}
corresponding to introducing $\vec x$ dependence into Eq. \eqref{PhiPsi} above.
Substituting into Eq. \eqref{gxy}, we get
\begin{equation}\label{gxy0}
0=-2(1+\Phi)\partial_x\Psi\partial_y\Psi - (1+\Phi)(1-\Psi) \partial_x\partial_y \Psi
+(1-\Psi)^2 \partial_x\partial_y \Phi +(1-\Psi)
(\partial_x \Phi\partial_y \Psi+ \partial_y\Phi\partial_x \Psi)~~~.
\end{equation}
To leading linear order in $\Phi$ and $\Psi$, this reduces to
\begin{equation}\label{leading}
0=\partial_x\partial_y[\Phi(t,\vec x)-\Psi(t, \vec x)]~~~,
\end{equation}
which assuming a smooth limit to vanishing spatial dependence gives the leading order result
\begin{equation}\label{leading1}
\Phi(t)=\Psi(t)~~~.
\end{equation}

We have not found a model-independent way to go beyond leading order, so we proceed by introducing
a specific model, from which we abstract a relation which we then use with no further reference
to the model.  Our model is to assume that $\Phi$ and $\Psi$ are dominated by a single plane wave
so that
\begin{align}
\Phi(t, \vec x) =&\Phi(t) e^{i\vec k \cdot \vec x}~~~,\cr
\Psi(t, \vec x) =&\Psi(t) e^{i\vec k \cdot \vec x}~~~.\cr
\end{align}
 With this assumption $\partial_x\equiv ik_x\,,\,\partial_y\equiv ik_y$,  which can then be factored from all terms of Eq. \eqref{gxy0}.  This leaves a purely algebraic relation between $\Phi(t)$ and $\Psi(t)$, which
can be reduced to the simple form
\begin{equation}\label{PhPsrel}
\Phi(t)=\frac{\Psi(t)}{1-2\Psi(t)}~~~,
\end{equation}
which to first order reproduces Eq. \eqref{leading1}.  Reexpressing Eq. \eqref{PhPsrel} in terms
of $\theta(t)$ and $\alpha(t)$ we find
\begin{align}\label{thetalpheq}
\alpha(t)=& \frac{\theta(t)}{2\theta(t)-1}~~~,\cr
\theta(t)=& \frac{\alpha(t)}{2\alpha(t)-1}~~~,\cr
2\theta(t)\alpha(t)=&\theta(t)+\alpha(t)~~~.\cr
\end{align}
These equations give the relation between $\alpha$ and $\psi$  that we will use in our analysis
beyond leading order.  It has the nice feature that  $\alpha(t)\equiv 1$ implies that  $\theta(t)\equiv 1$,
$\psi(t)\equiv a(t)$, which is the same property found for Eqs. \eqref{00comp} and \eqref{spatcomp} above. Thus our model relating $\alpha$ and $\psi$ is structurally compatible with the Einstein equations derived in the preceding section.

\subsection{Differential equation for $\theta$, reduction to the linearized case, and large and small
time behavior}

Since study of the linearized case in Sec. 3 shows that $\Psi(t)=1-\theta(t)$ is slowly varying in
comparison to the FRW expansion factor $a(t)$, we turn Eq. \eqref{spatcomp} into an equation
for $\theta$ by substituting $\psi(t)=a(t) \theta(t)/\theta(0)$.  Setting the pressure term $p$ to zero,
using the zeroth order FRW equation $2 \ddot a/a + (\dot a/a)^2=\Lambda$,  changing to the
dimensionless variable $x$ of Eq. \eqref{dimtime}
and using $a^\prime/a=(2/3) {\rm coth}(x)$ with
${}^\prime$ denoting $d/dx$, we arrive at
\begin{align}\label{thetdiffeq}
\theta^{\prime\prime}=&F[\theta^\prime,\theta]~~~,\cr
F[\theta^\prime,\theta]=&\frac{2}{3}\alpha^2[1-f+f/\alpha^4] \theta -
\frac{ (2\theta+1)}{2(2\theta-1)}\frac{(\theta^\prime)^2}{\theta} -\frac{4}{3} {\rm coth}(x) \frac{3\theta-1}{2\theta-1} \theta^\prime -\frac{2}{3}\theta~~~,\cr
\end{align}
with $\alpha$ in the first term of $F$ related to $\theta$ by $\alpha=\theta/(2\theta-1)$.  To check
that this reduces to the linearized case, we set $\theta=1-\Psi$ and keep only terms of first order
in $\Psi$, giving after a little algebra
\begin{equation}\label{firstPsi}
\Psi^{\prime\prime}+\frac{8}{3} {\rm coth}(x) \Psi^\prime=\frac{4}{3}(2f-1) \Psi ~~~,
\end{equation}
in agreement (using the perturbative relation $\Psi=\Phi$) with Eq. \eqref{evolution2}  above.
We note that since $\theta=1$ corresponds to $\alpha=1$, Eq. \eqref{thetdiffeq} implies that $F(0,1)=0$,
with the consequence that initial data $\theta^\prime(0)=0$ and $\theta(0)=1$ propagate forward in
time unchanged.

Since Eq. \eqref{thetdiffeq} is even in $x$ it admits $\theta$ to be an even function of $x$,
and so at $x=0$ there will be a regular solution with $\theta(0)$ an arbitrary constant, and
$\theta^{\prime}(0)=0$.  Making the ansatz that $\theta$ becomes infinite for large $x$,
Eq. \eqref{thetdiffeq} reduces to an equation with constant coefficients. Substituting
$\theta(x)=e^{\lambda x}$ we find the algebraic equation
\begin{equation}\label{lameq}
\lambda^2+\frac{4}{3} \lambda +\frac{4}{9} = \frac{1}{9}(15 f+1)~~~,
\end{equation}
with roots
\begin{equation}\label{roots}
\lambda=-\frac{2}{3} \pm \frac{1}{3}(15 f+1)^{1/2}~~~,
\end{equation}
to be compared with
\begin{equation}\label{linexp}
\lambda = -(2/3)[2 \pm (6f+1)^{1/2}]
\end{equation}
 found from the linearized equation in Sec. 3.
When $f=1$ Eq. \eqref{roots} gives an exponentially growing solution with $\lambda=2/3\simeq 0.67$, somewhat larger than the exponent $(2/3)(\surd 7-2)\simeq 0.43$ given for $f=1$ by Eq. \eqref{linexp}.  For the nonperturbative formula of Eq. \eqref{roots} the exponentially growing solution appears for $f>1/5$, whereas for the linearized
result of Eq. \eqref{linexp} the exponentially growing solution first appears for the larger value $f>1/2$.

\subsection{Numerical solution}

To solve for $\theta$ numerically, we employ stepwise forward integration of Eq. \eqref{thetdiffeq}
starting from $x=0$,
\begin{align}\label{stepwise}
\theta^\prime(i+1)=&\,\theta^\prime(i)+F[\theta^\prime(i),\theta(i)]\Delta x~~~,\cr
\theta(i+1)=&\,\theta(i)+ \theta^\prime(i) \Delta x~~~,\cr
\end{align}
with $\Delta x=1/800$, and then as a check with $\Delta x=1/1600$.  These two calculations differ
only in the 5th decimal place.  In Table II we give results with $f=1$ for $\theta$, and for the normalized ratio $\Psi(x)/\Psi(0)= [1-\theta(x)]/[1-\theta(0)]$, starting from $\theta(0)=1.1$, or equivalently $\Psi(0)=-0.1$, and $\theta^\prime(0)=0$.    The normalized ratio $\Psi(x)/\Psi(0)$ is directly comparable to $\hat \Phi$ calculated  in Sec. 3 and given in Table I;
 in the final column of Table II we give $\hat \Phi$ recalculated by the stepwise forward integration method used here for solving the $\theta$ differential equation, which agrees to within one or two parts per thousand with the iterative integral equation method used in Sec. 3.\footnote{For example, interpolating
 into Table II we find for z=0.54 that  $\hat \Phi=0.6 \times 1.095+ 0.4 \times 1.080 = 1.089$, compared
 with 1.089 in Table I.}   We see that the
final two columns of Table I agree to within a couple of percent, indicating that the linear perturbation
results of Sec. 3 suffice at present for phenomenological applications.

We have also checked
that with the iteration of Eq. \eqref{stepwise}, initial data $\theta(0)=1$ and $\theta^{\prime}(0)=0$
propagate forward in time unchanged.  For $\theta(0)>1$, the numerical solution for $\theta(x)$ with $f=1$ increases with increasing $x$, while for $\theta(0)<1$ the numerical solution for $\theta(x)$ decreases
with increasing $x$, corresponding respectively to a universe that expands more
rapidly, or more slowly, than the standard FRW solution $a(t)$.

\begin{table} [ht]
\caption{Values of $\theta(x)$ and $\Psi(x)/\Psi(0)$ calculated for $\theta(0)=1.1$, and in the final column the corresponding linearized equation result $\hat \Phi(x)$, versus $x$ and redshift $z$, taking  $f=1$, that is, with all of
the observed cosmological constant arising from a scale invariant but frame dependent action.}
\centering
\begin{tabular}{c  c c c c }
\hline\hline
$~z~$ & $~~~~x~~$& $~~~~~\theta(x)~~~$  & ~~~$\Psi(x)/\Psi(0)$~~~ &$ ~~~ \hat \Phi(x)~~~$ \\
\hline
0  & 1.169  &1.126      & 1.264   & 1.243      \\
0.1  & 1.054  & 1.122   &1.215    & 1.198        \\
 0.2 & 0.955  & 1.118   & 1.177   & 1.163        \\
0.3  & 0.868  & 1.115   & 1.146   & 1.135        \\
0.4 & 0.792   &  1.112   & 1.122  & 1.113        \\
0.5  & 0.726  &  1.110  & 1.103   & 1.095        \\
0.6  & 0.668  & 1.109   & 1.087   & 1.080        \\
0.7  & 0.616  & 1.107   & 1.074   & 1.069        \\
0.8  & 0.571  & 1.106   & 1.064   & 1.059        \\
0.9  & 0.530  & 1.106   & 1.055   & 1.051        \\
1.0  & 0.494  & 1.105   & 1.048   & 1.044        \\
\hline\hline
\end{tabular}
\label{tab2}
\end{table}

\subsection{Correspondence of the non-perturbative and perturbative solutions}

To conclude this section, we see that a nonperturbative treatment of the problem that we addressed by linearized perturbation theory in Sec. 3 leads again to a one-parameter family of cosmologies, parameterized now by the initial value $\theta(0)$.  In the perturbative treatment, the corresponding parameter was $\Phi(0)$, and this parameter is related to the one used in the nonperturbative analysis  by $\Phi(0)\simeq \Psi(0)=1-\theta(0)$.  For
$\theta(0)=1 ~,~~\Phi(0)=0$, the nonperturbative and perturbative evolutions both have as the solution
the standard FRW expansion factor $a(t)$; that is, $\theta(t)$ remains equal to 1 (and $\Phi(t)$ remains 0) for all times $t$.  For $\theta(0)>1~,~~\Phi(0)<0$ the $f=1$ nonperturbative and pertubative evolutions both give a universe that expands faster than the standard FRW solution $a(t)$, as discussed further below, and for $\theta(0)<1~,~~\Phi(0)>0$ the $f=1$ nonperturbative and pertubative evolutions both
give a universe that expands more slowly than the standard FRW solution $a(t)$.   There is thus a direct
qualitative correspondence between the two methods of treating the model of Eq. \eqref{ansatz}, and for  parameter values $\theta(0)$ close to unity, there is only a
small quantitative difference.   We have also seen that the nonperturbative treatment gives a relatively
straightforward way of rederiving the pertubation equation of Eq. \eqref{evolution2}.

\section{Cosmographic equations and application to the ``Hubble tension''}

To discuss cosmography we rewrite the  metric in terms of proper time $\tau$ as in Eq. \eqref{metric1},
\begin{equation}\label{rewrite}
ds^2=d\tau^2-\psi^2[\tau] d\vec x^{\,2} ~~~,
\end{equation}
which we recall takes
the standard  FRW form with $\tau$ replacing $t$ and with $\psi[\tau]\equiv \psi(t)=a(t)\theta(t)/\theta(0)$ replacing $a(t)$.  Thus in comparisons with experiment, it is $\tau$ that will be the true physical time,
and $\psi[\tau]$ the true expansion rate.  We proceed now with a number of applications of the analysis of the previous sections.  Where we give perturbative results, we use the first order accurate equality
$\Phi(t)\simeq \Psi(t)$, which is equivalent to  $\theta(t)\alpha(t) \simeq 1$.

\begin{itemize}
\item Substituting the quadratic approximation to $\hat \Phi$ from Table I, we get
a simple approximate formula for the transformation from coordinate time $t$ to proper time $\tau$,
\begin{align}\label{explicttime}
\tau(t)=&\int_0^t \alpha(u)du= \int_0^t du \alpha(0)[\alpha(u)/\alpha(0)]
\simeq \alpha(0) \int_0^t du [1+\Phi(0)(\hat\Phi(u)-\hat \Phi(0))] \cr
&\simeq \alpha(0) \int_0^t du [1+\Phi(0) C (u/t_0)^2]
=\alpha(0)t [1+ (\Phi(0)C/3)(t/t_0)^2]~~~,\cr
\end{align}
with $C=0.244$.
This has the inversion
\begin{equation}\label{explicitinversion}
t[\tau]=(\tau/\alpha(0)) [1- (\Phi(0)C/3)(\tau/\tau_0)^2]~~~.
\end{equation}
\item
 We now have to find the transformation that relates  $H_0$ and $t_0$ with the physical quantities $H_0^{\rm Pl}$ and $\tau_0^{\rm Pl}$ measured by Planck \cite{planck} (and earlier, by WMAP \cite{wmap}) from the  CMB.
  We note from Table I that for
redshifts $z$ greater than 100, and therefore
at and before decoupling,  $\Phi(t)  $ is very accurately the constant $\Phi(0)$, and
thus $\alpha(t)$ is very accurately equal to $\alpha(0)$ .  So the
transformation from coordinate time $t$ to proper time $\tau$ is just a constant rescaling
$\tau=\alpha(0) t$. Since to first order $\theta(t) \simeq 1/\alpha(t)$, for large redshifts $\theta(t) \simeq \theta(0)$, and thus   $\psi[\tau]= a(t)\theta(t)/\theta(0) \simeq a(t)=a(\tau/\alpha(0))$.   Hence the relation
between $H_0^{\rm Pl}$ and $H_0$ is  just a rescaling by $\alpha(0)$,
\begin{equation}\label{rescaling}
H_0^{\rm Pl} = H_0/\alpha(0)~,~~~H_0= H_0^{\rm Pl} \alpha(0)~~~,
\end{equation}
and introducing the age of the universe $\tau_0^{\rm Pl}$ as measured by Planck, we have
\begin{align}\label{rescaling1}
\tau_0^{\rm Pl} = &H_0t_0/H_0^{\rm Pl}=t_0 \alpha(0)~~~,\cr
H_0t_0=&H_0^{\rm Pl}\tau_0^{\rm Pl}=\frac{2x_0}{3\surd{\Omega_\Lambda}}=0.946~~~.\cr
\end{align}
We will see that the effect of these rescalings is to make all physical process depend on
$\hat \Phi(t)$ through the combination $\hat \Phi(t)-\hat \Phi(0) = C(t/t_0)^2$, so that if
the late time increase coefficient $C$ were set to zero, there would be no physical effects.
This is essential, because with $C=0$ the relation between the coordinate time $t$ and the proper
time $\tau$ of Eq. \eqref{explicttime} becomes a constant rescaling for all times, and a change
in the units in which time is measured should have no effect on physical consequences of the equations.

\item
The nonperturbative generalizations of the standard cosmological distance measures are obtained by
making the substitution $a(t) \to \psi[\tau] = \psi(t)=a(t) \theta(t)/
\theta(0) $   in the usual formulas, with the following results, where
the subscript $0$ denotes  the present time, and $1$ denotes the past time at which an object at coordinate distance $r_1$ emitted a signal.

  For waves of wavelength $\lambda$ and frequency $\nu$, we have
\begin{equation}\label{redshift}
1+z_{\rm eff}=\frac{\psi[\tau_0]}{\psi[\tau_1]} = \frac{1}{\psi[\tau_1]} = \frac{\theta(0)}{a(t_1)\theta(t_1)}~~~.
\end{equation}
where in the second  equality we used $\psi[\tau_0]=1$.

 The parallax distance distance $d_P$, luminosity distance $d_L$, angular diameter distance $d_A$,
and proper motion or comoving angular diameter distance $d_M$, are given by
\begin{align}\label{distmeas}
d_P=&r_1\psi(t_0)~~~,\cr
d_L=&r_1\psi(t_0)(1+z_{\rm eff})~~~,\cr
d_A=&r_1\psi(t_1)~~~,\cr
d_M=&r_1\psi(t_0)~~~,
\end{align}
with ratios
\begin{align}\label{ratios}
\frac{d_A}{d_L}=&\frac{1}{(1+z_{\rm eff})^2}~~~,\cr
\frac{d_M}{d_L}=&\frac{1}{1+z_{\rm eff}}~~~,\cr
\frac{d_A}{d_P}=&\frac{1}{1+z_{\rm eff}}~~~.\cr
\end{align}

\item  The perturbative formulas needed to evaluate the above expressions are:\footnote{Whenever multiplied by $\Phi(0)$, which is first order in magnitude, other quantities can be taken to
    zeroth order.  Hence $\Phi(0)x_\tau^2\simeq \Phi(0)x^2$,  etc.}

\begin{align}\label{pertcosmog}
\frac{\theta(0)}{\theta(t)}=&1+\Phi(0)[\hat{\Phi}(t)-\hat{\Phi}(0)]\cr
=&1+\Phi(0)C x_\tau^2/x_0^2.~~~,\cr
a(t) =& \left(\frac{\Omega_m}{\Omega_{\Lambda}}\right)^{1/3} \left(\sinh\Big(x_\tau [1-\frac{1}{3}\Phi(0)C \frac{x_\tau^2}{x_0^2}]
\Big)\right)^{2/3}~~~,\cr
x_\tau=&\frac{3}{2} \surd{\Omega_\Lambda}H_0^{\rm Pl} \tau~~~.\cr
\end{align}

\item As noted in Sec. 2,
 the effective Hubble parameter is defined as
the proper time derivative
\begin{equation}\label{hubble1}
H_{\rm eff}[\tau]= \frac{d\psi[\tau]/d\tau}{\psi[\tau]}~~~,
\end{equation}
or in terms of $t$,
\begin{equation}\label{hubble2}
H_{\rm eff}(t)=\frac{d\psi(t)/dt}{\alpha(t)\psi(t)}=\frac{d\big(\theta(t)a(t)\big)/dt}{\alpha(t)\theta(t)a(t)}
=\frac{1}{\alpha(t)}\left(\frac{\dot a(t)}{a(t)}+\frac {\dot \theta(t)}{\theta(t)}\right)
=\frac{1}{\alpha(t)}\left(H(t)+\frac{\dot \theta(t)}{\theta(t)}\right)~~~.
\end{equation}
In the linearized perturbation limit this is (using $\alpha \theta\simeq 1$ )
\begin{equation}\label{hubble3}
H_{\rm eff}(t)\simeq [1-\Phi(t)]H(t) - \dot \Phi(t)~~~,
\end{equation}
From this and Eq. \eqref{hubbleapp}, working to first order accuracy, we find at the present proper time $\tau_0$ that
\begin{equation}\label{hubble4}
\frac{H_{\rm eff}[\tau_0]}{H_0^{\rm Pl}}\simeq 1-\Phi(0)[ (\hat\Phi(x_0)-1)(1+\frac{3}{2}\Omega_m)+ \frac{3}{2} \surd \Omega_{\Lambda}
d\hat \Phi/dx|_{x=x_0}]\simeq 1-0.867 \Phi(0)~~~,
\end{equation}
with $(\hat \Phi(x_0)-1)(1+(3/2)\Omega_m)$  contributing $0.3615$  and with $\frac{3}{2} \surd \Omega_{\Lambda}
d\hat \Phi/dx|_{x=x_0}$ contributing $0.5055$ to the coefficient $0.867$. If we use the quadratic
approximation to $\hat \Phi(x)$ given in the final column of Table I, Eq. \eqref{hubble4} becomes
\begin{equation}\label{hubble5}
\frac{H_{\rm eff}[\tau_0]}{H_0^{\rm Pl}}\simeq 1-\Phi(0)C(1+\frac{3}{2}\Omega_m+3\surd{\Omega_\Lambda}/x_0)
=1-0.877\Phi(0)~~~,
\end{equation}
in good agreement with Eq. \eqref{hubble4}.

\item
Much recent attention has been paid to a tension between values of the local Hubble constant directly measured from redshifts using supernova samples calibrated with Cepheids \cite{riess}, or supernova
samples calibrated with the tip of the red giant branch \cite{freedman}, versus the value at early cosmic times inferred from the study of fluctuations in the CMB radiation \cite{planck}, \cite{knox}.  This tension may require recalibration of the distance ladder used in direct redshift measurements \cite{feeney}, or may indicate a need for new physics \cite{mortsell}.  To interpret the Hubble tension in terms of new physics via a frame-dependent dark energy, we take $f=1$ and use Eq. \eqref{hubble4} to  fit the ratio $H_{\rm local}/H_0^{\rm Pl}$, as summarized in Table III, where we give the parameter
$\Phi(0)$  needed to fit the ratio of the local Hubble constant value of \cite{riess} or \cite{freedman} to the CMB Hubble value quoted in \cite{knox}.   This then gives a prediction of our model for the Hubble constant value extracted from baryon acoustic oscillations (BAO) by the method of  Alam et al. \cite{boss}, as discussed in Appendix B, with the result given in the fourth column of Table III.   The prediction coming from the Cepheid determination of the local Hubble constant \cite{riess} is $4.8 \pm 2.1$ percent  larger than
the BAO value.  As argued in \cite{lemos}, this tension is a reflection of the fact that our model is
a late time model, which generically cannot accommodate the Hubble values from both \cite{riess} and \cite{boss}.

\item  When $1+\Phi(0)=\alpha(0)<1$, the expansion rate in our model is enhanced over that in the standard
  FRW cosmology for all times.  To see this, take the $\tau$ derivative of the logarithm of $\psi[\tau] = \theta(t(\tau)) a(t(\tau))/\theta(0)$,
converting derivatives on the right to $t$ derivatives using $d/d\tau = (1/\alpha(t)) d/dt$, to get
\begin{equation}\label{expanrate}
\frac{d \log(\psi)}{d\tau}= \frac{1}{\alpha(t)} \left[ \frac{d \log(a(t))}{dt} + \frac{d \log (\theta(t))}{dt}\right] ~~~,
\end{equation}
with $t$ on the right hand side understood to be $t(\tau)$.
Since for $\alpha(0) <1$  we have $\frac{1}{\alpha(t)}>1 $ and $d \theta(t)/dt>0$,
Eq. \eqref{expanrate} implies that
\begin{equation}\label{expanrate1}
\frac{d \log(\psi)}{d\tau}-\frac{d \log(a)}{dt}>0~~~;
\end{equation}
that is, in our model with $f=1$ and negative $\Phi(0)$,  the expansion rate of the universe
at all times is greater than in standard FRW cosmology.

\item  Corresponding to the fact that expansion rate of the universe is altered in our model for
$\Phi(0)\neq 0$, the age of the universe is changed.  This can be calculated, to leading order in
perturbations, as follows.  We first note that just as the age $t_0$ of the FRW universe is fixed by the
requirement $a(t_0)=1$, in similar fashion the proper time age of the universe $\tau_0$ in our model is fixed\footnote{To elaborate, the age of the universe is the elapsed proper time between redshift zero and redshift infinity.  When $\psi=1$, the effective redshift
$z_{\rm eff}$ vanishes, and at $\tau=0$ the effective redshift is infinite since $\psi[0]=a(0)=0$.}
by the condition
$\psi[\tau_0]=1$.  Writing $t(\tau_0)=t_0 +\Delta t$ we get from $\psi(t)=a(t)\theta(t)/\theta(0)$ the equation
\begin{align}
1=&\psi[\tau_0]=a(t(\tau_0))\theta(t(\tau_0))/\theta(0)\cr
=&a(t_0+\Delta t)[1-\Phi(0)(\hat \Phi(t_0+\Delta t)-\hat \Phi(0))] \cr
\simeq & 1+ \Delta t \frac{da}{dt}(t_0)-\Phi(0)(\hat \Phi(t_0)-\hat \Phi(0))  ~~~,\cr
\end{align}
which can be solved to give $\Delta t$,
\begin{equation}\label{deltatsoln}
\Delta t=\frac{\Phi(0)(\hat \Phi(t_0)-\hat \Phi(0))}{\frac{da}{dt}(t_0)}=\frac{\Phi(0)(\hat \Phi(t_0)-\hat \Phi(0))}{H_0}\simeq \frac{\Phi(0)(\hat \Phi(t_0)-1)}{H_0^{\rm Pl}}~~~.
\end{equation}
From the definition of the proper time in Eq. \eqref{propertime} we get
\begin{align}\label{tauzero}
\tau_0=& \int_0^{t(\tau_0)} du \alpha(u) \cr
=&\int_0^{t_0+\Delta t} du [1+\Phi(0) \hat \Phi(u)] \cr
=&t_0+\Delta t+ \Phi(0) \int_0^{t_0} du \hat \Phi(u)~~~,\cr
\end{align}
and remembering that $t_0= \tau_0^{\rm Pl}/\alpha(0)\simeq  \tau_0^{\rm Pl} (1-\Phi(0))$
this  gives to first order
\begin{align}\label{difftaut}
\tau_0-\tau_0^{\rm Pl}=&-\Phi(0)\tau_0^{\rm Pl}+\Delta t + \Phi(0) \int_0^{t_0} du \hat \Phi(u) \cr
\simeq &\frac{\Phi(0)}{H_0^{\rm Pl}} \left[ \hat \Phi(t_0)-1
+ \frac{2}{3\surd \Omega_\Lambda} \int_0^{x_0} dx\hat \Phi(x) -\tau_0^{\rm Pl} H_0^{\rm Pl}\right] ~~~.
\end{align}
Using the approximation $\hat \Phi(x) \simeq 1+ C(x/x_0)^2$ this gives
\begin{equation}\label{finaldifft}
\tau_0-\tau_0^{\rm Pl}= \frac{\Phi(0)C}{H_0^{\rm Pl}}\left(1+\frac{2 x_0}{9 \surd{\Omega_{\Lambda}}}\right)\simeq \frac{1.315\Phi(0)C}{H_0^{\rm Pl}}~~~,
\end{equation}
which is used to calculate the final column of Table III, using $\tau_0^{\rm Pl}=13.83 {\rm Gyr}$, and $1/H_0^{\rm Pl}=\tau_0^{\rm Pl}/0.946=14.62 {\rm Gyr}$.
\end{itemize}

\begin{table} [ht]
\caption{Results of fits of $\Phi(0)$ to the Hubble tension obtained by the Cepheid \cite{riess} and red giant \cite{freedman} methods. The second column gives the ratio of the local Hubble measurement to the CMB  Hubble value $ H_0^{\rm Pl} =67.27\, {\rm km}\, {\rm s}^{-1}\, {\rm Mpc}^{-1} $ quoted in the review \cite{knox}.   On the Cepheid line \cite{riess}, the error  is statistical. On the red giant line \cite{freedman}, the
first error is statistical and the second is systematic. The third column gives the value of $
\Phi(0)$ needed to fit the second column using Eq. \eqref{hubble4}.   The fourth column gives the
ratio of the Hubble parameter to the corresponding baryon acoustic oscillation (BAO) value \cite{boss}  at  $z_{\rm eff}=0.51$, as calculated in Appendix B.  The fifth column gives the correction to the age of the
universe from Eq. \eqref{finaldifft}.}
\centering
\begin{tabular}{c  c c  c c }
\hline\hline
Method  &$\frac{H_{\rm local}}{H_0^{\rm Pl}}$    & $\Phi(0)$   &$\frac{H_{\rm eff,\,0.51}}{H_{{\rm BAO},\,0.51}}$ & $\tau_0-\tau_0^{\rm Pl}$({\rm Gyr})\\
\hline
Cepheid &$1.100 \pm 0.023$       &$-0.114\pm 0.026$       &$ 1.048\pm 0.021$   &    $-0.53\pm 0.12$ \\
red giant &~$1.038\pm 0.015 \pm0.025$ ~       & ~$-0.043\pm0.017\pm0.028$ ~     &~ $1.018\pm 0.019 \pm 0.012$ ~ &~ $-0.20\pm0.08\pm0.13$~\\
\hline\hline
\end{tabular}
\label{tab3}
\end{table}

\vfill
\eject

To summarize, a scale invariant but frame dependent dark energy can enhance the local
Hubble constant value  without spoiling the excellent CMB angular fits, with a universe that
is expanding faster and hence is younger than is suggested by the standard FRW cosmology. However,
our model, in which changes from the standard FRW cosmology occur only at late time, tends to give too large a value for the Hubble constant extracted by the BAO method.  We look forward to future experiments to give an enlarged and improved data set against which to give a definitive test of our model.

\section{Acknowledgements}
I wish to thank Juan Maldacena, Paul Steinhardt, and especially  Matias Zaldarriaga for helpful conversations.  I also thank Angelo Bassi and Stephen Boughn for reading the initial version of this paper.  This work was done in part at the Aspen Center for Physics, which is supported by National Science
Foundation grant PHY-1607611.

\appendix

\section{Perturbation equations rewritten in terms of $\Phi$ and $\Psi$}

In \cite{adler4} we gave the metric perturbation equations for $f \neq 0$ in $B=0$ gauge in terms of $A$, $E$, and $F$, taken as functions of both $t$ and $\vec x$.  Here we give the same equations when rewritten in terms of $\Psi$, $\Phi$, and $F$, obtained by substituting Eq.  \eqref{EAelim} and algebraic
simplification.\footnote{We remind the reader that here we follow the notation of \cite{adler4} and
\cite{wein} and omit the normalization factor $1/\theta(0)$.  This has no effect on the final result of
Eq. \eqref{final2}, which is independent of the normalization of $a$, even though Eqs. \eqref{ijeq}-\eqref{00eq} are sensitive to the normalization of $a$ when spatial dependences are included.}

The $ij$ equation takes the form
\begin{align}\label{ijeq}
0=&\delta_{ij}X+\partial_i\partial_j Y~~~,\cr
X=&2[a \ddot{a} + 2(\dot a)^2] \Phi + a \dot a \dot \Phi + 6 a \dot a \dot \Psi +a^2 \ddot \Psi -
\nabla^2 \Psi   \cr
+&4 \pi G a^2(\delta \rho -\delta p -\nabla^2 \pi^S)-4\pi G[\dot a a^2 4(p+\rho)+ a^3(\dot \rho/3
+\dot p)] F \cr
+& \Lambda f a^2 (t_{00}/2-E)~~~,\cr
Y=&8\pi G a^2 \pi^S+\Phi-\Psi~~~.\cr
\end{align}
The $i0$ equation becomes
\begin{equation}\label{i0eq}
0=  -2 (\dot a  /a) \partial_i \Phi -2 \partial_i \dot \Psi   -8 \pi G(p+\rho)(\partial_i \delta u -a \partial_i F)+ \Lambda f t_{0i}~~~,
\end{equation}
and the $00$ equation becomes\footnote{Under the gauge changes $\Delta_g(a^2 A)=2 a \dot a\epsilon_0$,  $\Delta_g E=2 \dot \epsilon_0$, $\Delta_g(aF)=-\epsilon_0$, $\Delta_g\delta \rho=\dot \rho \epsilon_0$, $\Delta_g \delta p=\dot p \epsilon_0$, $\Delta_g\pi^S=0$, $\Delta_g  \delta u=-\epsilon_0$, $\Delta_g \Phi=\Delta_g \Psi=0$, all terms of Eqs. \eqref{EAelim} and \eqref{ijeq}--\eqref{00eq} are invariant {\it except} the $\Lambda f$ terms, reflecting the fact that the effective action of Eq. \eqref{effact3} breaks four-space
    general coordinate invariance.  When only three-space general coordinate transformations are considered (which is the most general invariance of Eq. \eqref{ansatz}), on can take $\epsilon_0=0$, which
    is consistent with our then setting $F(t)=0$.}
\begin{align}\label{00eq}
0=&Z=-(1/a^2)\nabla^2 \Phi-3 (\dot a/a)\dot \Phi - 6(\ddot a/a) \Phi -3 \ddot \Psi -6 (\dot a/a) \dot \Psi  \cr
+&4 \pi G[\delta \rho + 3 \delta p + \nabla^2 \pi^S + a(\dot \rho+ 3 \dot p) F]\cr
+&\Lambda f (t_{00}/2 + 3 E) ~~~.\cr
\end{align}
Taking the linear combination $(1/4)a^2Z+(3/4)X+(1/4)\nabla^2 Y$ gives
\begin{align}\label{combo}
0=&- \nabla^2\Psi +3 a \dot a \dot \Psi +3 (\dot a)^2 \Phi + \Lambda f a^2  t_{00}/2 \cr
+&4 \pi G a^2[\delta \rho-3 \dot a (p+\rho)F]~~~,\cr
\end{align}
and using this to eliminate $-\nabla^2\Psi$ from $X$ we get
\begin{align}\label{final1}
a^2 \ddot \Psi+[2 a \ddot a + (\dot a)^2] \Phi+a \dot a \dot \Phi + 3 a \dot a \dot \Psi
=&4\pi G a^2 (\delta p + \nabla^2 \pi^S)+2\Lambda f a^2 [\Phi-\dot a F -a \dot F]\cr
+&4 \pi G a^3 \dot p F~~~.\cr
\end{align}
In the limit that the metric perturbations are functions only of $t$, using $F(t)=0$ and $\nabla^2 \pi^S(t)=0$ this simplifies, after division by $a^2$, to
\begin{equation}\label{final2}
\ddot \Psi+[2  (\ddot a/a) + (\dot a/a)^2] \Phi+ (\dot a/a) \dot \Phi + 3  (\dot a/a) \dot \Psi
=4\pi G  \delta p +2\Lambda f \Phi~~~.
\end{equation}
When $\Psi=\Phi$, this gives Eq. \eqref{evolution1} of the text.  The point of the manipulations leading
to Eq. \eqref{final2} is to eliminate both energy densities $\delta \rho$ and $t_{00}$ from the equation used to solve for $\Phi$.

\section{Comparison at $z_{\rm eff}=0.51$ of the Hubble parameter predicted by our model with the value inferred from BAO measurements}

Baryon acoustic oscillation  measurements using three clusters of galaxies grouped around redshifts
 of 0.38, 0.51, and 0.61 give \cite{boss} a Hubble constant  $H_0^{\rm BAO} =67.3\pm 1.0   \,{\rm km}\, {\rm s}^{-1}\, {\rm Mpc}^{-1}$, nearly identical to the CMB value \cite{knox} of   $ H_0^{\rm Pl} =67.27\pm 0.60\, {\rm km}\, {\rm s}^{-1}\, {\rm Mpc}^{-1} $.  To compare this with the prediction of our
 model, we calculate the ratio of the Hubble expansion parameters $H(t)$ evaluated at $t$ corresponding to the central redshift 0.51.  For the Hubble expansion parameter implied by the BAO measurements, we use the FRW cosmology formula of Eq. \eqref{hoft} with $H_0$ replaced by the BAO
 measured value, giving
\begin{align}\label{hoft1}
H_{{\rm BAO},\,0.51}=H_0^{\rm BAO}[\Omega_m(1.51)^3+\Omega_{\Lambda}]^{1/2}\cr
=&89.9\pm 1.3\, {\rm km}\, {\rm s}^{-1}\, {\rm Mpc}^{-1}~~~.\cr
\end{align}
For the Hubble expansion parameter implied by our model, we use Eqs. \eqref{redshift} and \eqref{pertcosmog} to determine the $x_\tau$ value corresponding to $1+z_{\rm eff}=1.51$, and then
substitute this into Eq. \eqref{hubble3}, which we expand to first order accuracy,
\begin{align}\label{hubbleapp}
H_{\rm eff}(t)\simeq &[1-\Phi(t)]H(t) - \dot \Phi(t)\cr
\simeq & H_0^{\rm  Pl}\surd{\Omega_\Lambda}\left\{\coth\left[x_\tau\left(1-\frac{\Phi(0)C}{3}\frac{x_\tau^2}{x_0^2}\right)\right]
-\frac{\Phi(0)C}{x_0^2}(x_\tau^2 \coth(x_\tau) + 3 x_{\tau})\right\}~~~.\cr
\end{align}
This yields  
\begin{align}\label{hoft2}
H_{\rm eff,\,0.51}=&94.2\pm1.3\,{\rm km}\, {\rm s}^{-1}\, {\rm Mpc}^{-1}~~{\rm for}~~ \Phi(0)=-0.114 \pm 0.026~~~,\cr
H_{\rm eff,\,0.51}=&91.5\pm1.0\pm1.1\,{\rm km}\, {\rm s}^{-1}\, {\rm Mpc}^{-1}~~{\rm for}~~\Phi(0)=-0.043\pm0.017\pm0.028~~~,\cr
\end{align}

giving the numbers used to form the ratio $\frac{H_{\rm eff,\,0.51}}{H_{{\rm BAO},\,0.51}}$ in Table III.


\begin{thebibliography}{99}
\bibitem{mathematica} I received this notebook from  Fethi Ramazano\v glu (who got it from Hans
Bantilan) and used it to compute all algebraic Einstein equations and covariant conservation equations in what follows.
\end{thebibliography}
\end{document}